\documentstyle[preprint,aps]{revtex}
\begin{document}
\preprint{MZ-TH/95-02}
\draft
\load{}{\it}
\load{\Large}{\it}
\load{\LARGE}{\it}

\title{Triangle Diagram with Off-Shell Coulomb T-Matrix
for (In-)Elastic Atomic and Nuclear Three-Body Processes}

\author{\bf E. O. Alt$^1$, A. S. Kadyrov$^2$,
A. M. Mukhamedzhanov$^{2,3}$, and M. Rauh$^1$}

\address{$^1$ Institut f\"ur Physik, Universit\"at Mainz,
D-55099 Mainz, Germany\\
$^2$ Institute for Nuclear Physics, Ulugbek, Tashkent 702132,
Uzbekistan \\
$^{3}$ Present address: Cyclotron Institute, Texas A\&M
University, College Station, TX 77843, USA}

\maketitle
\begin{abstract}
The driving terms in three-body theories of elastic and
inelastic scattering of a charged particle off a bound state
of two other charged particles contain the fully off-shell
two-body Coulomb T-matrix describing the intermediate-state
Coulomb scattering of the projectile with each of the charged
target particles.
Up to now the latter is usually replaced by the Coulomb
potential, either when using the multiple-scattering
approach or when solving three-body integral equations.
General properties of the exact and the approximate on-shell
driving terms are discussed, and the accuracy of this
approximation is investigated numerically, both for atomic and
nuclear processes including bound-state excitation, for
energies below and above the corresponding three-body
dissociation threshold, over the whole range of scattering
angles.
\end{abstract}

\vspace{15mm}

Classification Numbers: 3480B, 3410, 2410, 2510

\newpage
\section{Introduction}

Elastic and inelastic scattering of a particle off a
two-particle cluster
is conveniently formulated in terms of the exact
three-body integral equations in momentum space (Faddeev 1960)
which, after suitable manipulations, can be given the
practical form of multichannel equations of the two-body
Lippmann-Schwinger type (Alt {\em et al} 1967). If the
impinging, and one or both of the target particles, are
charged, modifications of this theory are required
(Veselova 1971,1978, Alt 1978, Alt {\em et al} 1978,1980).
Nevertheless, the basic, effective-two-body, multi-channel
structure of the equations can be preserved.

An alternative approach for calculating the (in-)elastic
scattering amplitude, which is well-suited especially
for high energies, is based on the multiple-scattering
expansion of the corresponding three-body transition operator.
Obviously, there exists a close correspondence between the
terms occurring here and the various contributions to the
effective potential of the integral equations approach.

In applications of either formalism to charged-particle
scattering, terms occur which contain the fully off-shell
two-particle Coulomb T-matrix, describing the
intermediate-state Coulomb scattering of various charged
two-body subsystems. It is evident that the singularity
structure of the latter in momentum space renders the
calculation of such terms a rather
difficult task. Hence, in numerical work (see, e. g., Alt {\em
et al} 1985, Alt {\em et al} 1994, and references therein) the
Coulomb T-matrix is frequently replaced by its Born
approximation, namely the Coulomb potential, which drastically
reduces the analytical and numerical effort required. But the
reliability of such an approximation, in the following called
`Coulomb-Born approximation', is difficult to estimate.

The first numerical studies of the quality of such a
Coulomb-Born approximation appear to have been performed by
Sinfailam and Chen (1972), who calculated the lowest-order
multiple-scattering contributions to the elastic scattering
amplitude for various atomic three-body processes, with the
full two-body Coulomb T-matrix and with the latter replaced
by the Coulomb potential. It turned out that, for most of the
three-body systems studied, and for the energy and
angular ranges considered, the Coulomb-Born approximation
failed rather dramatically, the more in fact the lower the
energy was. Of course, if the latter is sufficiently high,
both the exact and the approximate expression are expected
to eventually yield the same answer; this was, indeed, found
although at an energy larger than expected. This investigation,
however, concerned only a limited variation of the scattering
angle and the (altogether rather high) three-body energy.
In particular, energies in the neighbourhood of the
dissociation threshold, which represent critical tests of
the reliability of the calculational procedures since the
essential singularity of the two-body Coulomb T-matrix at
zero subsystem energy plays a particularly important role
there, were not considered at all.

Theoretical and numerical studies of the lowest-order
(off-shell) driving terms, which occur in the three-body
approach, have been performed by van Haeringen (1979), and Kok
{\em et al} (1979,1980,1981,1982). However, the investigations
were restricted to negative total three-body energies, i.e.,
to energies below the composite-particle breakup threshold (and
to the case that the masses of all three particles are equal, and
the charges of the two particles involved in the
intermediate-state Coulomb scattering are of equal sign).
In some situations, which relate more to nuclear reactions
with light nuclei, the Coulomb-Born approximation was found to
be reasonably accurate. But in other cases, which are more
typical for the situation prevalent in atomic physics, its lack of
accuracy considerably diminishes its usefulness, similarly to
the situation found at high energies.

In this paper we address ourselves to a systematic
investigation of the quality of the Coulomb-Born approximation,
by comparing the exact with the corresponding approximate
on-shell driving terms for elastic and inelastic scattering
in three-body systems. In particular, we study the dependence on
several important parameters. Namely, energies
are considered from the reaction threshold up to such high
values that the Coulomb-Born approximation practically yields
the exact result. Scattering angles are varied over the whole
angle regime, with particular emphasis on very small angles
which are favoured in higher-energy scattering.
Furthermore, different masses of the two particles which
experience the intermediate-state Coulomb scattering, are
allowed for: two light, one light and one heavy, and two heavy
particles. Also the case that the magnitude of the charge of
one of these two particles is larger than one is considered.
And, finally, we also study the collisional excitation
of the bound system. However, in the present investigation
we restrict ourselves to repulsive intermediate-state
Coulomb interactions.

In Sect. \ref{three} we briefly sketch the three-body
approach to elastic and inelastic scattering and, in
particular,
introduce the relevant driving terms. Some known results for
the two-particle off-shell Coulomb amplitude which are used in
the subsequent investigations are collected in Section
\ref{tcoul}. In Sect. \ref{gen}, general properties of these
elastic and inelastic driving terms are discussed, both in
their exact form as well as if the intermediate-state Coulomb
T-matrix is replaced by the Coulomb potential. Numerical tests
of the accuracy of this approximation are presented for
several typical atomic and nuclear three-body systems
in Sect. \ref{appl}, both for energies below and
above the corresponding bound state dissociation thresholds.
The results obtained are summarized in Sect. \ref{disc}.

Natural units $\hbar = c = 1$ are chosen. Furthermore, the
conventional notation for two-body quantities: $A_{\alpha}
\equiv A_{\beta \gamma}$, with $\alpha \ne \beta \ne \gamma
\ne \alpha$, is adopted.

\newpage
\section{Three-particle model of elastic and inelastic
scattering}
\label{three}

Consider three particles with masses $m_{\nu}$ and charges
$e_{\nu}$, $\nu = 1, 2, 3$. We are interested in the
reaction $\alpha+(\beta \gamma)_m \to \alpha+ (\beta
\gamma)_n$, where particle $\alpha$, having center-of-mass
momentum ${\rm{\bf{q}}}_{\alpha}$, impinges on the bound state
of particles $\beta$ and $\gamma$; the bound state
wave function, belonging to the binding energy
$\hat E_{\alpha m}$ (with quantum numbers $m$), is
denoted by $|\psi_{\alpha m}\rangle$.
In the final state, the bound state quantum
numbers are characterized by the index $n$, and the
center-of-mass momentum of particle $\alpha$ is
${\rm{\bf{q}}}_{\alpha}'$.

The corresponding (in-)elastic scattering amplitude
${\cal T}_{\alpha n, \alpha m}({\rm{\bf{q}}}_{\alpha}',
{\rm{\bf{q}}}_{\alpha}; E)$ can be
written as matrix element of an effective-two-body operator
${\cal T}_{\alpha n, \alpha m}(E + i0)$ between the plane wave
states describing the free asymptotic motion of the clusters,
\begin{equation}
{\cal T}_{\alpha n, \alpha m}({\rm{\bf{q}}}_{\alpha}',
{\rm{\bf{q}}}_{\alpha}; E) =
\langle{\rm{\bf{q}}}_{\alpha}'|{\cal T}_{\alpha n, \alpha m}
(E + i0)|{\rm{\bf{q}}}_{\alpha}\rangle, \label{amp}
\end{equation}
with the total three-body energy $E$ being connected with the
incoming and outgoing momenta by the on-shell condition
\begin{equation}
E =\frac{q_{\alpha}^{2}}{2 M_{\alpha}} + \hat E_{\alpha m} =
\frac{q_{\alpha}'^{2}}{2 M_{\alpha}} + \hat E_{\alpha n}.
\label{oes}
\end{equation}
Here, $M_{\alpha} = m_{\alpha}(m_{\beta} + m_{\gamma})/
(m_{\alpha} + m_{\beta} + m_{\gamma})$ is the $\alpha$-channel
reduced mass.

According to the effective-two-body formulation of the
three-body scattering theory (Alt {\em et al} 1967), the
operators ${\cal T}_{\alpha n, \alpha m}(z)$, together
with the corresponding operators ${\cal T}_{\beta n,
\alpha m}(z)$ appropriate for rearrangement
scattering, are given as solutions of the following set of
coupled Lippmann-Schwinger-type equations
\begin{equation}
{\cal T}_{\beta n, \alpha m} (z) ={\cal V}_{\beta n, \alpha m}
(z) + \sum_{\gamma=1}^{3} \sum_{rs}{\cal V}_{\beta n, \gamma
r}(z) {\cal G}_{0;\gamma,rs}(z) {\cal T}_{\gamma s, \alpha m}(z).
\label{ls}
\end{equation}
The sums over r and s extends at least over all bound states
of subsystem $\gamma$, consisting of particles $\alpha$ and
$\beta$. Thus, the solution of (\ref{ls}) yields
simultaneously the transition amplitudes for all two-fragment
reactions. We note in parentheses that in the presence of
Coulomb forces, the standard methods of integral equations
theory can not be applied directly to (\ref{ls}) to obtain
the physical reaction amplitudes, due to lack of compactness
of their kernel (for a proper procedure in this case see
Alt (1978), Alt {\em et al} (1978,1980)).

In leading order in an iterative solution of eq. (\ref{ls}),
the on-shell amplitudes describing elastic and inelastic
processes are given by the on-shell `diagonal' effective
potential
\begin{eqnarray}
{\cal T}_{\alpha n, \alpha m}({\rm{\bf{q}}}_{\alpha}',
{\rm{\bf{q}}}_{\alpha}; E) & \approx &
{\cal V}_{\alpha n, \alpha m}({\rm{\bf{q}}}_{\alpha}',
{\rm{\bf{q}}}_{\alpha}; E) \nonumber \\
&=& \langle{\rm{\bf{q}}}_{\alpha}'|\langle\psi_{\alpha n}|
U_{\alpha \alpha}'(E + i0)|\psi_{\alpha m}\rangle
|{\rm{\bf{q}}}_{\alpha}\rangle. \label{born}
\end{eqnarray}
The auxiliary three-body operators $U_{\beta \alpha}'$
satisfy the equations
\begin{equation}
U_{\beta \alpha}'(z) = \bar\delta_{\beta \alpha}
G_0^{-1}(z) + \sum_{\gamma = 1}^{3} \bar\delta_{\beta \gamma}
T_{\gamma}'(z) G_0(z) U_{\gamma \alpha}'(z). \label{uprime}
\end{equation}
Here, $G_0(z) = (z - H_0)^{-1}$ denotes the free three-body
resolvent, and $\bar\delta_{\beta \alpha} = 1 -
\delta_{\beta \alpha}$ the anti-Kronecker symbol.
$T_{\gamma}'(z)$ is the sum of the two-particle
off-shell Coulomb transition operator $T_{\gamma}^C(z)$ plus
a `remainder short-range transition operator' in subsystem
$\gamma$. For the definition of the effective free Green
function ${\cal G}_{0;\gamma,rs} (z)$ and further details
we refer to Alt (1978), Alt {\em et al} (1978). We only
mention that the nondiagonal effective-potential parts
${\cal V}_{\beta n, \alpha m} (z)$  with $\beta \neq \alpha$,
which are the driving terms relevant for rearrangement
scattering, are defined similarly to (\ref{born}) in terms
of the operators $U_{\beta \alpha}'(z)$. They
are, however, not considered in the present investigation.

A Neumann series expansion of (\ref{uprime}) leads to the
following representation of the `diagonal' on-shell effective
potentials
\begin{eqnarray}
{\cal V}_{\alpha n, \alpha m}({\rm{\bf{q}}}_{\alpha}',
{\rm{\bf{q}}}_{\alpha}; E) &=& \sum_{\gamma = 1}^{3}
\bar\delta_{\gamma \alpha} \langle{\rm{\bf{q}}}_{\alpha}'|
\langle\psi_{\alpha n}|T^C_{\gamma}(E + i0)
|\psi_{\alpha m}\rangle|{\rm{\bf{q}}}_{\alpha} \rangle +
\cdots. \label{effpot}
\end{eqnarray}
Explicitly shown are here only the leading Coulomb
contributions due to (Coulomb) single-scattering in the
intermediate state, which will be dealt with in the following
investigation. The dots indicate single-scattering
contributions from the remainder
short-range transition operator as well as higher-order
rescattering terms. We note in parentheses that, if only
two of the three particles are charged,
and if the shorter-ranged interactions are represented in
purely separable form, the infinite series
for ${\cal V}_{\alpha n, \alpha m}({\rm{\bf{q}}}_{\alpha}',
{\rm{\bf{q}}}_{\alpha}; E)$ collapses exactly to just
the terms written down explicitly in (\ref{effpot}).
Furthermore,
the terms (\ref{effpot}) would occur as the Coulomb
single-rescattering contributions to the (in-)elastic
scattering amplitude in a multiple-scattering approach.

Let us introduce for the terms on the right-hand side
of (\ref{effpot}) the notation
\begin{eqnarray}
{\cal V}^{T^C}_{\gamma, n m}
({\rm{\bf{q}}}'_{\alpha}, {\rm{\bf{q}}}_{\alpha}; E):\, &=&
\langle{\rm{\bf{q}}}'_{\alpha}|\langle\psi_{\alpha n}|
T^C_{\gamma}(E + i0 )|\psi_{\alpha m}\rangle
|{\rm{\bf{q}}}_{\alpha}\rangle \nonumber\\
&=& \int \frac{d^3k}{(2 \pi)^3} \psi^*_{\alpha n}
({\rm{\bf{p}}}'_{\alpha}) \,\hat T^C_{\gamma}\!\!
\left({\rm{\bf{p}}}'_{\gamma},{\rm{\bf{p}}}_{\gamma}; E+ i0 -
\frac{k^2}{2 M_{\gamma}}\right)
\psi_{\alpha m}({\rm{\bf{p}}}_{\alpha}). \qquad \label{telg}
\end{eqnarray}
The various subsystem momenta are defined as
\begin{eqnarray}
{\rm{\bf{p}}}_{\alpha} &=& \epsilon_{\alpha \gamma}
\left({\rm{\bf{k}}} + \frac{\mu_{\alpha}}{m_{\beta}}
{\rm{\bf{q}}}_{\alpha}\right), \qquad
{\rm{\bf{p}}}'_{\alpha} = \epsilon_{\alpha \gamma}
\left({\rm{\bf{k}}} + \frac{\mu_{\alpha}}{m_{\beta}}
{\rm{\bf{q}}}'_{\alpha}\right), \nonumber \\
{\rm{\bf{p}}}_{\gamma} &=& \epsilon_{\gamma \alpha}
\left({\rm{\bf{q}}}_{\alpha} +
\frac{\mu_{\gamma}}{m_{\beta}}{\rm{\bf{k}}}\right), \qquad
{\rm{\bf{p}}}'_{\gamma} = \epsilon_{\gamma \alpha}
\left({\rm{\bf{q}}}'_{\alpha} + \frac{\mu_{\gamma}}{m_{\beta}}
{\rm{\bf{k}}}\right). \label{mom}
\end{eqnarray}
Here, $\mu_{\alpha} = m_{\beta}m_{\gamma}/
(m_{\beta} + m_{\gamma})$ is the reduced mass of the pair
$(\beta \gamma)$. For convenience, the antisymmetric symbol
$\epsilon_{\beta \alpha} = - \epsilon_{\alpha \beta}$,
with $\epsilon_{\alpha \beta} = +1$ if
$(\alpha,\beta)$ is a cyclic ordering of (1,2,3), is used.
Moreover, the Coulomb T-matrix when read in the two-particle
space is characterized by a hat, $\hat T^C$. The
graphical representation of ${\cal V}^{T^C}_{\gamma, n m}$
is given in Fig. 1.

Similarly we define the quantities
${\cal V}^{V^C}_{\gamma, n m}({\rm{\bf{q}}}'_{\alpha},
{\rm{\bf{q}}}_{\alpha})$ by the expression (\ref{telg})
but with the Coulomb T-matrix $T^C_{\gamma}$ replaced by the
Coulomb potential $V^C_{\gamma}$. They will be referred to
as the `Coulomb-Born approximation'. We mention
that for simple types of bound state wave functions
${\cal V}^{V^C}_{\gamma, n m}({\rm{\bf{q}}}'_{\alpha},
{\rm{\bf{q}}}_{\alpha})$ can be calculated analytically
(Lewis 1956).

\newpage
\section{The two-particle Coulomb T-matrix} \label{tcoul}

In the literature there exist several equivalent expressions
for the fully off-shell two-body Coulomb T-matrix in momentum
space. A fairly complete account of results concerning
two-body Coulomb scattering can be found in van Haeringen
(1985). We will take the relevant formulae from van Haeringen
(1979), and Kok and van Haeringen (1980) (see also van
Haeringen and van Wageningen (1975), Chen and Chen (1972)),
which were developed
there with regard to their applicability for the evaluation of
expressions like the effective potential (\ref{telg}). Recall
that we confine ourselves in the present investigation to
case that the two particles participating in the
intermediate-state rescattering have charges of equal sign.

The following T-matrix representation is used,
\begin{equation}
\hat T^C_{\gamma}({\rm{\bf{p}}}',{\rm{\bf{p}}};
\hat E_{\gamma}+i0) = V^C_{\gamma}({\rm{\bf{\Delta}}})
\left \{1 + \frac{I(y)}{x} \right \}, \label{tc1}
\end{equation}
with
\begin{eqnarray}
V^C_{\gamma}({\rm{\bf{\Delta}}}) &=& \frac{4 \pi
e_{\alpha}e_{\beta}}{\Delta^2}, \label{vc}\\
I(y) &=& F_{i \eta_{\gamma}}(y) - F_{i \eta_{\gamma}}(y^{-1}).
\label{vc1}
\end{eqnarray}
Here, ${\rm{\bf{\Delta}}}={\rm{\bf{p}}}'-{\rm{\bf{p}}}$ is
the momentum transfer; $\hat E_{\gamma}$ is the
center-of-mass energy of the interacting pair of particles
$\alpha$ and $\beta$, and $\eta_{\gamma} =
e_{\alpha}e_{\beta}\sqrt{\mu_{\gamma}/2 \hat E_{\gamma}}$
the appropriate Coulomb parameter. Moreover, we have
introduced the abbreviations
\begin{eqnarray}
x^2 &=& 1+\frac{(p'^{2} - 2 \mu_{\gamma} \hat E_{\gamma})
(p^{2} - 2 \mu_{\gamma} \hat E_{\gamma})}
{2 \mu_{\gamma} \hat E_{\gamma} \Delta^2}, \label{x} \\
y &=& \frac{x + 1}{x - 1}, \label{y}\\
F_{i \eta_{\gamma}}(y) &\equiv&_2F_1(1, i \eta_{\gamma}; 1 +
i \eta_{\gamma}; y), \label{aux}
\end{eqnarray}
with $_2F_1(a,b;c;y)$ being the hypergeometric function.
Note that in writing down the formulae of this section we
always assume $\hat E_{\gamma} \neq 0$ (formulae suitable
for $\hat E_{\gamma}$ in the vicinity of zero can be found
in van Haeringen (1979)). For $ \Re\, i \eta_{\gamma} > 0$
the integral representation for the hypergeometric function
is particularly useful,
\begin{equation}
F_{i \eta_{\gamma}}(y) = i \eta_{\gamma} \int_0^1 dt \,
t^{i \eta_{\gamma} - 1} (1 - yt)^{-1}, \quad
\Re \, i \eta_{\gamma} > 0. \label{ir}
\end{equation}
It leads to the following expression for the Coulomb T-matrix
\begin{eqnarray}
\hat T^C_{\gamma}({\rm{\bf{p}}}',{\rm{\bf{p}}};
\hat E_{\gamma}+i0) =
V^C_{\gamma}({\rm{\bf{\Delta}}}) \left\{ 1- 4i \eta_{\gamma}
\int_0^1 dt \frac{t^{i \eta_{\gamma}}}{4t -(x^2-1)
(1-t)^2}\right\}, \quad \Re \,i \eta_{\gamma} > 0 .
\label{tir}
\end{eqnarray}
Furthermore, for $\Re \, x < 0$, which implies $|y| < 1$, one
has the series representation for the function $I(y)$:
\begin{equation}
I(y) = 1 - (-y)^{i \eta_{\gamma}}
\frac{\pi \eta_{\gamma}}{\sinh \pi \eta_{\gamma}} +
2 \eta_{\gamma}^2 \sum_{n = 1}^{\infty} \frac{y^n}
{n^2 + \eta_{\gamma}^2}. \label{series}
\end{equation}

We mention that off the energy shell, in the limit of
vanishing momentum transfer, the following behaviour is
easily derived:
\begin{equation}
\frac{I(y)}{x} \stackrel{\Delta \to 0}{\sim}
\Delta \{c_1 + c_2 \ln \Delta  + c_3 \Delta +
o\,(\Delta) \},\label{mt0}
\end{equation}
that is, the most singular part of the repulsive Coulomb
T-matrix is given by the potential, for all (off-shell)
subsystem energies $\hat E_{\gamma}$. This leads for the ratio
${\cal R}=\hat T^C_{\gamma}({\rm{\bf{p}}}',{\rm{\bf{p}}};
\hat E_{\gamma}+i0)/V^C_{\gamma}({\rm{\bf{\Delta}}})$
to the well-known result
\begin{equation}
|{\cal R}|\stackrel{\Delta \to 0}{=} 1 + O(\Delta)
\quad \forall \quad E_{\gamma}. \label{mt1}
\end{equation}
Of course, on the energy shell one has $|{\cal R}|=1$ for
all scattering angles and energies.

Kok and van Haeringen (1980) (see also van Haeringen and
Kok (1984), van Haeringen (1985)) have derived useful general
bounds on ${\cal R}$. Introduce the notation $\cos \vartheta =
{\rm{\bf{p}}}' \cdot {\rm{\bf{p}}}/pp'$. For instance,
they have proved the following inequalities valid for
a repulsive Coulomb interaction:
\begin{eqnarray}
0 \leq {\cal R} \leq 1, \quad \forall \quad p,p',\cos
\vartheta \quad \mbox{for} \quad \hat E_{\gamma} < 0,
\quad e_{\alpha}e_{\beta}>0, \label{r1} \\
0 \leq |{\cal R}| \leq 1, \quad \forall \quad p,p', \cos
\vartheta \quad \mbox{for} \quad \hat E_{\gamma} > 0,
\quad e_{\alpha}e_{\beta}>0, \label{r2}
\end{eqnarray}
which contain (\ref{mt1}) as special case. Illustrative
graphical representations for the ratio ${\cal R}$
are given in Kok and van Haeringen (1980).

\newpage
\section{General properties of ${\cal V}^{T^C}_{\gamma,
\lowercase{n m}}$ and ${\cal V}^{V^C}_{\gamma,
\lowercase{n m}}$} \label{gen}

In this section we will discuss some general properties of the
lowest-order contributions to the elastic effective potential,
graphically represented by Fig. 1, and investigate the effects
which arise from the replacement of the Coulomb T-matrix by
the Coulomb potential.

Introducing the Coulomb T-matrix representation (\ref{tc1})
into expression (\ref{telg}) one sees that the Coulomb potential
can be taken out of the integral. Hence we can write
\begin{eqnarray}
{\cal V}^{T^C}_{\gamma, n m}({\rm{\bf{q}}}'_{\alpha},
{\rm{\bf{q}}}_{\alpha}; E) =
V^C_{\gamma}({\rm{\bf{\Delta}}}_{\alpha})
{F}^{T^C}_{\gamma, n m}({\rm{\bf{q}}}'_{\alpha},
{\rm{\bf{q}}}_{\alpha}; E), \label{telg1}
\end{eqnarray}
where
\begin{eqnarray}
{F}^{T^C}_{\gamma, n m}({\rm{\bf{q}}}'_{\alpha},
{\rm{\bf{q}}}_{\alpha}; E) =
\int \frac{d^3k}{(2 \pi)^3} \psi^*_{\alpha n}
({\rm{\bf{p}}}'_{\alpha}) \left \{1 + \frac{I(y)}{x} \right \}
\psi_{\alpha m}({\rm{\bf{p}}}_{\alpha}), \label{iel1}
\end{eqnarray}
with
\begin{eqnarray}
x^2 &=& 1+\frac{\left[p^2_{\gamma}-2\mu_{\gamma}(E + i0-
k^2/2 M_{\gamma})\right] \left[p_{\gamma}'^{2} - 2
\mu_{\gamma}
(E + i0 - k^2/2 M_{\gamma})\right]}
{2 \mu_{\gamma}(E + i0 -k^2/2 M_{\gamma}) \,
\Delta^2_{\alpha}},
\label{xx}\\
{\rm{\bf{\Delta}}}_{\alpha} &=&\epsilon_{\gamma \alpha}
({\rm{\bf{q}}}'_{\alpha} -{\rm{\bf{q}}}_{\alpha}).
\end{eqnarray}
$y$ is defined as before, cf. eq. (\ref{y}), and
$V^C_{\gamma}$ by (\ref{vc}). Similarly,
${\cal V}^{V^C}_{\gamma, n m}$ can be written as
\begin{equation}
{\cal V}^{V^C}_{\gamma, n m}({\rm{\bf{q}}}'_{\alpha},
{\rm{\bf{q}}}_{\alpha}) =
V^C_{\gamma}({\rm{\bf{\Delta}}}_{\alpha})
{F}^{V^C}_{\gamma, n m}({\rm{\bf{q}}}'_{\alpha},
{\rm{\bf{q}}}_{\alpha}), \label{velg}
\end{equation}
with
\begin{eqnarray}
{F}^{V^C}_{\gamma, n m}({\rm{\bf{q}}}'_{\alpha},
{\rm{\bf{q}}}_{\alpha}) &=& \int\frac{d^3k}{(2 \pi)^3}
\psi_{\alpha n}^*({\rm{\bf{p}}}'_{\alpha})
\psi_{\alpha m}({\rm{\bf{p}}}_{\alpha}) \nonumber \\
&=& \int \frac{d^3k}{(2 \pi)^3} \psi_{\alpha n}^*
({\rm{\bf{k}}}) \psi_{\alpha m}({\rm{\bf{k}}} + \mu_{\alpha}
{\rm{\bf{\Delta}}}_{\alpha} /m_{\beta}) \label{iel2}
\end{eqnarray}
for $m=n$ being called bound state form factor, and for
$m \neq n$ transition form factor. In the following
investigation we will mainly be concerned with the ratio
\begin{eqnarray}
R_{\gamma,n m}({\rm{\bf{q}}}'_{\alpha},{\rm{\bf{q}}}_{\alpha};
 E) : = \frac{{\cal V}^{T^C}_{\gamma, n m}
({\rm{\bf{q}}}'_{\alpha},{\rm{\bf{q}}}_{\alpha}; E)}
{{\cal V}^{V^C}_{\gamma, n m}({\rm{\bf{q}}}'_{\alpha},
{\rm{\bf{q}}}_{\alpha})} = \frac{F^{T^C}_{\gamma, n m}
({\rm{\bf{q}}}'_{\alpha},{\rm{\bf{q}}}_{\alpha}; E)}
{F^{V^C}_{\gamma, n m}({\rm{\bf{q}}}'_{\alpha}
{\rm{\bf{q}}}_{\alpha})}. \label{rel}
\end{eqnarray}

Let us state some general results for $R_{\gamma,n m}$.

(i) For elastic scattering, both ${\cal V}^{T^C}_{\gamma,mm}$
and ${\cal V}^{V^C}_{\gamma, m m}$ diverge for
${\rm{\bf{q}}}'_{\alpha}-{\rm{\bf{q}}}_{\alpha} \rightarrow 0$
because of (\ref{tc1}) and (\ref{vc}); however, as eq.
(\ref{tc1}) with (\ref{mt0}) implies, their ratio tends in
magnitude towards the value one,
\begin{equation}
\left \vert R_{\gamma,m m}({\rm{\bf{q}}}'_{\alpha},
{\rm{\bf{q}}}_{\alpha}; E)\right \vert \rightarrow 1
\quad{\rm for} \quad {\rm{\bf{q}}}'_{\alpha} \to
{\rm{\bf{q}}}_{\alpha}\quad \forall \quad E. \label{rel1}
\end{equation}

(ii) The inequalities (\ref{r1}) and (\ref{r2}) result in
the following inequalities for the elastic ratio
$R_{\gamma,0 0}$, where the index 0 denotes the ground state:
\begin{eqnarray}
0 < R_{\gamma,0 0}({\rm{\bf{q}}}'_{\alpha},
{\rm{\bf{q}}}_{\alpha};E)
\leq 1, \quad \mbox{for} \quad E<0, \label{rmm1} \\
0 < |R_{\gamma,0 0}({\rm{\bf{q}}}'_{\alpha},
{\rm{\bf{q}}}_{\alpha}; E)|\leq 1, \quad \mbox{for} \quad E>0.
\label{rmm2}
\end{eqnarray}
That is, for elastic scattering off a target in the ground
state the Coulomb-Born approximation always {\em overestimates}
the exact driving term. Or, in other words, the error made by
approximating in (\ref{telg}) the two-body Coulomb T-matrix
by the Coulomb potential is of known sign. Note that no
analogous bounds result if either one or both bound state
wave functions have nodes.

(iii) Since for large two-body subsystem energies $T_{\gamma}^C$
approaches the Born approximation $V_{\gamma}^C$, we expect
for elastic and inelastic scattering
\begin{equation}
R_{\gamma,n m}({\rm{\bf{q}}}'_{\alpha},{\rm{\bf{q}}}_{\alpha};E)
\stackrel{E \to \infty}{\longrightarrow} 1. \label{he}
\end{equation}
However, it is obvious that for (\ref{he}) to hold the energy
$E$ must be higher than that for which on the two-body level
we have $T_{\gamma}^C(\hat E_{\gamma}) \approx V_{\gamma}^C$.
For, in ${\cal V}^{T^C}_{\gamma, m m}({\rm{\bf{q}}}'_{\alpha},
{\rm{\bf{q}}}_{\alpha}; E)$ the Coulomb T-matrix
enters for all $\gamma$-subsystem energies from $E$ down
to minus infinity, $- \infty < \hat E_{\gamma} =
E -k^2/2 M_{\gamma} \leq E$. Thus a behaviour like (\ref{he})
can result only as a combined effect of $\hat T_{\gamma}^C(E -
k^2/2 M_{\gamma})$ being approximately equal to
$V_{\gamma}^C$ over the whole range of momenta
${\rm{\bf{k}}}$ for which the momentum-space bound state
wave functions differ appreciably from zero.

(iv) For excitation we deduce from (\ref{iel2})
\begin{equation}
\lim_{{\rm{\bf{q}}}'_{\alpha} \to
{\rm{\bf{q}}}_{\alpha}} {F}^{V^C}_{\gamma, n m}
({\rm{\bf{q}}}'_{\alpha}, {\rm{\bf{q}}}_{\alpha}) = 0
\quad \mbox{for} \quad n \neq m \label{ort}
\end{equation}
because of the orthogonality of the bound state wave
functions. Of course, on the energy shell (\ref{oes}) the
momentum transfer can never vanish. However, in forward
direction its magnitude, though remaining non-zero, can
become small, thereby making $R_{\gamma,n
m}({\rm{\bf{q}}}'_{\alpha},{\rm{\bf{q}}}_{\alpha};E)$
rather large. Furthermore, since the
inequalities (\ref{r1}) and (\ref{r2}) do not
result in inequalities like (\ref{rmm1}) or (\ref{rmm2}) for
the ratio $R_{\gamma,n m}$, values smaller and
larger than one are possible.

\newpage
\section{Numerical results} \label{appl}

When performing the integration over the magnitude of the
momentum ${\rm{\bf{k}}}$ of the noninteracting particle
$\gamma$ in the exact driving term (\ref{telg}) we must, for
a given three-body energy $E>0$, treat the two regions
$2 M_{\gamma}E< k^2 <\infty $ (region I) and $k^2< 2
M_{\gamma}E$ (region II) which correspond to negative and
positive subsystem energy $\hat E_{\gamma} = E -k^2/2 M_{\gamma}$,
differently. In region II, choosing $\Re \, x < 0$, the
representation (\ref{series}) for the two-body Coulomb T-matrix
can be used. In order to select an expression suitable for region
I we first note that the Coulomb parameter $\eta_{\gamma}$
occurring in $\hat T^C_{\gamma}$ depends on the
integration variable ${\rm{\bf{k}}}$ via
\begin{equation}
\eta_{\gamma} = e_{\alpha}e_{\beta}\mu_{\gamma}/
\sqrt{2 \mu_{\gamma} (E+ i0 -k^2/ 2 M_{\gamma})}. \label{somm}
\end{equation}
Thus, $\eta_{\gamma}$ is imaginary.
Since, as mentioned at the beginning, we restrict
ourselves to that part of the elastic effective potential
which contains Coulomb rescattering between the projectile
and the target particle whose charge is of equal sign,
i.e., $e_{\alpha}e_{\beta} > 0$, in region I the condition
${\Re} \,i \eta_{\gamma} > 0$ is always fulfilled so that the
representation (\ref{tir}) is applicable. Of course, if
expression (\ref{telg}) is calculated for a negative
three-body energy $E < 0$, only region I is relevant.

In order to show that the numerical quadratures can be
performed reliably
even for positive three-body energies (where the integration
contour touches the essential singularity of the Coulomb
T-matrix at zero subsystem energy), we present in Fig. 2
as a typical example the real and the imaginary part of
(\ref{telg}), with all integrations except the one over
the magnitude of ${\rm{\bf{k}}}$ having been performed,
as function of $k$. We choose the elastic reaction $H(e,e)H$
to be discussed below, for a center-of-mass projectile
energy of 100 eV and a scattering angle of 20 degree.
Inspection reveals that the real part of this integrand
and its first derivative are very smooth in the neighbourhood
of that momentum for which the subsystem energy $E -k^2/2
M_{\gamma} = 0$, in spite of the fact that, as discussed
above,
on either side of this point a completely different expression
for the two-body Coulomb T-matrix is used. A similarly smooth
behaviour is observed for the imaginary part. For other
energies and scattering angles, and for other projectiles,
qualitatively similar results are obtained (or course, if
the projectile energy is smaller than the magnitude of the
binding energy of the incoming bound state, which corresponds
to negative total three-body energy, the imaginary
part of the integrand is identically zero). For comparison
we also include the analogous (real) integrand of the
Coulomb-Born approximation (\ref{velg}) which, as is to be
expected from (\ref{r2}), is so much larger in magnitude
than the real part of the exact expression that the absolute
value of the ratio (\ref{rel}) is smaller than one, in
accordance with (\ref{rmm2}).
We furthermore mention that it was carefully checked that
the driving term (\ref{telg}) behaved smoothly and
yielded the same value when the ionisation threshold was
approached from below and from above.

\subsection{Atomic reactions}

For definiteness we consider only reactions
where the bound states in the initial and in the final state
are described by hydrogen-like wave functions. Let us begin by
assuming the bound states to be in their ground states before
and after the collision, i. e., we take $m = n = \{100\}$
(with the usual notation $\{n \ell m \}$ for the set of
hydrogenic quantum numbers). Thus
\begin{equation}
\psi_{\alpha m}({\rm{\bf{p}}}) \equiv \psi_{100}({\rm{\bf{p}}})
= \frac{8 \pi^{1/2} \kappa_{\alpha 0}^{5/2}}
{(p^2 + \kappa_{\alpha 0}^2)^2}, \label{psi}
\end{equation}
and the same expression for the outgoing bound state. Here,
$\kappa_{\alpha 0}^2=- 2 \mu_{\alpha} \hat E_{\alpha 0}=
(\mu_{\alpha} e_{\beta} e_{\gamma})^2$, $\hat E_{\alpha 0}$
being the Coulomb ground state energy of the bound pair
$(\beta \gamma)$. Introducing the wave functions (\ref{psi})
into expressions (\ref{telg}) and (\ref{velg}) for the exact
and the approximate driving terms, we can integrate
analytically over the
azimuthal angular integration variable. The remaining
three-dimensional integrals have to be done numerically.
It is to noted that for a few selected cases we have verified
that our results coincide with those of Sinfailam and Chen
(1972) within the accuracy with which the numbers can be
extracted from their figures.

In Fig. 3 we present the ratio
$\left \vert R_{ee,0 0}({\rm{\bf{q}}}'_{\alpha},
{\rm{\bf{q}}}_{\alpha}; E) \right \vert$ for
elastic scattering of electrons off hydrogen atoms in
their ground state, as function of the center-of-mass
scattering angle $\vartheta$ (with $\cos \vartheta
= {\rm{\bf{q}}}_{\alpha}
\cdot {\rm{\bf{q}}}_{\alpha}'/q_{\alpha} q_{\alpha}'$) and
of the center-of-mass projectile energy,
starting at the elastic threshold. Here and
in the following, the notation is such that the two
particles, which undergo Coulomb scattering in the
intermediate state and which are denoted by the subsystem
index $\gamma$ in (\ref{rel}), are explicitly indicated.
Furthermore, the ground state quantum numbers $\{100\}$
in the initial and the final state are abbreviated by the
index `0'. As is apparent, for each
energy $ \left \vert R_{ee,0 0}({\rm{\bf{q}}}'_{\alpha},
{\rm{\bf{q}}}_{\alpha}; E) \right \vert$
starts in the forward direction at the value one, thus
satisfying the condition (\ref{rel1}). For energies beyond
100 keV the ratio becomes practically equal to one again, in
accordance with (\ref{he}). Away from these regions this
figure allows us to estimate the error made when using the
Born approximation for the Coulomb T-matrix in the driving
term (\ref{telg}). Inspection reveals that over a wide range
of angles and energies the Coulomb-Born approximation
dramatically overestimates the exact rescattering
contribution. It is interesting to note that the minimum
value of $ \left \vert R_{ee,0 0}({\rm{\bf{q}}}'_{\alpha},
{\rm{\bf{q}}}_{\alpha}; E) \right \vert$ of approximately
0.27 is reached not in the backward direction but at an angle
of about 60 degree (for a projectile energy of around 70 eV).

The effect of increasing the mass of one of the particles
undergoing intermediate-state Coulomb scattering can be inferred
from Fig. 4. There we show the ratio $\left \vert R_{p e^+,00}
({\rm{\bf{q}}}'_{\alpha},{\rm{\bf{q}}}_{\alpha}; E)\right \vert$
for positron-hydrogen scattering, with the positron scattering
off the proton. It is apparent that the picture is similar to the
previous case, but minimum has deepened (to about 0.054 around
60 eV) and occurs in backward direction.

The case of two heavy particles experiencing Coulomb scattering
in the intermediate state is presented in Fig. 5, which contains
the ratio $\left \vert R_{p p,0 0}({\rm{\bf{q}}}'_{\alpha},
{\rm{\bf{q}}}_{\alpha}; E) \right \vert$ for proton-hydrogen
scattering. Evidently, the mass effect is very pronounced, the
Coulomb-Born approximation failing completely nearly everywhere,
even at extremely small scattering angles for higher energies.
The inadequacy of the Coulomb-Born approximation becomes still
more pronounced if one of the heavy particles has a charge
of magnitude greater than one. This can be inferred from
$\left \vert R_{p\,C^{6+},0 0}({\rm{\bf{q}}}'_{\alpha},
{\rm{\bf{q}}}_{\alpha}; E) \right \vert$ for the elastic
reaction $C^{6+} + H(1s)$, with $C^{6+}$ considered
as a structureless particle, which is displayed in Fig. 6.
In both cases the minimum (of the order $10^{-6}$ or less)
occurs at 180 degree.

In order to study the quality of the Coulomb-Born approximation
for excitation we calculated the exact and the approximate
driving terms (\ref{telg}) and (\ref{velg}) by assuming that
the outgoing bound state is in a 2s-state, characterized by a
wave function
\begin{equation}
\psi_{\alpha n}({\rm{\bf{p}}}) \equiv \psi_{200}({\rm{\bf{p}}})
= 64 (2 \pi \kappa_{\alpha 0}^{5})^{1/2} \frac{4p^2 -
\kappa_{\alpha 0}^2}{(4p^2 + \kappa_{\alpha 0}^2)^3}.
\label{psi1}
\end{equation}
To simplify the notation, the bound state quantum numbers
$\{200\}$ will be abbreviated by the index `1'.
We have already pointed out in Sect. \ref{gen} that for
excitation the magnitude of the ratio (\ref{rel})
is no longer bounded from above by the value one. In fact,
the Coulomb-Born approximation becomes very small in forward
direction, thereby giving rise to large values of the ratio.
This situation is exemplified in Fig. 7 where we have plotted
$\left \vert R_{ee,0 1}({\rm{\bf{q}}}'_{\alpha},
{\rm{\bf{q}}}_{\alpha}; E) \right \vert$ for
excitation of hydrogen atoms from the 1s- to the
2s-state by electron impact. It is seen to take
on values larger and smaller than one. The value one is
again reached at higher energies similar to those
for elastic scattering shown in Fig. 3. Furthermore,
although barely visible in the figure,
$\left \vert R_{ee,0 1}({\rm{\bf{q}}}'_{\alpha},
{\rm{\bf{q}}}_{\alpha}; E) \right \vert$ develops a
minimum of 0.27 at 45 eV projectile energy and 90 degree.
Interestingly, this minimum value coincides with that
for the elastic ratio and, as was the case there, occurs
at non-backward angles. Since, when approaching the forward
direction, $\left \vert R_{ee,0 1}({\rm{\bf{q}}}'_{\alpha},
{\rm{\bf{q}}}_{\alpha}; E) \right \vert$ increases sharply,
this figure has been cut off at a scattering angle of 1 degree.
It is clear that the larger the energy is, the relatively
smaller values of the momentum transfer are accessible and
thus the smaller the Coulomb-Born approximation becomes (cf.
eq. (\ref{ort})). Of course, for sufficiently high energy the
near-vanishing of ${\cal V}^{V^C}_{ee,0 1}
({\rm{\bf{q}}}'_{\alpha},{\rm{\bf{q}}}_{\alpha}; E)$ in the
forward direction has to be counterbalanced by a corresponding
near-vanishing of ${\cal V}^{T^C}_{ee,0 1}
({\rm{\bf{q}}}'_{\alpha}, {\rm{\bf{q}}}_{\alpha}; E)$, in
order that their ratio eventually approaches the value one.
We only mention that a similar picture is obtained for the
ratio $\left \vert R_{pp,0 1}({\rm{\bf{q}}}'_{\alpha},
{\rm{\bf{q}}}_{\alpha}; E) \right \vert$ for
excitation of hydrogen atoms from their ground to the
same excited state by proton impact.

Since due to these kinematic effects the ratio (\ref{rel})
is no longer an appropriate quantity to visualize the quality
of the Coulomb-Born approximation (except for scattering
angles larger than, say, 90 degree), we plot
in Fig. 8 only the absolute value of the exact driving term
$\left \vert {\cal V}^{T^C}_{ee,0 1}
({\rm{\bf{q}}}'_{\alpha}, {\rm{\bf{q}}}_{\alpha}; E) \right
\vert$ for excitation of hydrogen atoms by
electron impact, as function of the scattering angle and
the projectile energy. It is a very smooth quantity, however
ranging over 14 orders of magnitude in the range of
parameters considered. The mass effect is again illustrated
by considering the exact driving term $\left \vert
{\cal V}^{T^C}_{pp,0 1}({\rm{\bf{q}}}'_{\alpha},
{\rm{\bf{q}}}_{\alpha}; E) \right \vert$ for excitation of
hydrogen atoms by proton impact, displayed in  Fig. 9. Not
only is the range of accessible values smaller than in the
previous case; but there appears also to be
more structure in the amplitude. We point out that,
at small scattering angles, the exact driving terms for both
reactions increase in magnitude with increasing the energy,
and decrease only at rather high energies;
the energy where the maximum is reached lies the higher
the smaller the angle is.

\subsection{Nuclear reactions}

In order to investigate whether these rather large deviations
from the value one of $ \left \vert R_{\gamma,0 0} \right
\vert$ do hold also for other systems, we considered two
examples of elastic deuteron-nucleus scattering, i.e., one
of the three particles is a neutron. The nuclear interactions
are assumed to be given in separable form. Recall that under
such conditions, the expression (\ref{telg}), with $\gamma$
denoting the proton-nucleus subsystem, represents the
{\em exact} `diagonal' contribution to the on-shell effective
potential. The deuteron bound state wave function used here
is the same as that considered, e. g., by Alt {\em et al}
(1985) in their calculation of elastic proton-deuteron
scattering observables. That is, the deuteron is assumed to
be in an pure S state, with a momentum space wave function
\begin{equation}
\psi_{0}({\rm{\bf{p}}}) = \frac{N}{(p^2 + \kappa_d^2)
(p^2 + \beta^2)}. \label{yam}
\end{equation}
Here, $\kappa_d=\sqrt{- m_N \hat E_d}$, with $\hat E_d = -
2.226$ MeV being the deuteron binding energy and
$ m_N $ the nucleon mass; N is a normalisation constant, and
$\beta$ a parameter fitted to the low energy neutron-proton
scattering parameters in the $^3\!S_1$-state.

First we look at proton-deuteron scattering. With this bound
state wave function we have calculated the ratio
$\left \vert R_{pp,0 0}({\rm{\bf{q}}}'_{\alpha},
{\rm{\bf{q}}}_{\alpha}; E) \right \vert$ with proton-proton
intermediate-state Coulomb scattering, again as
function of the center-of-mass scattering angle and proton
bombarding energy. Note that the threshold for the deuteron
breakup is at a center-of-mass proton kinetic energy
of 2.226 MeV. The results are shown in Fig. 10.
Clearly, $\left \vert R_{pp,0 0}({\rm{\bf{q}}}'_{\alpha},
{\rm{\bf{q}}}_{\alpha}; E) \right \vert$ satisfies the general
constraints (\ref{rel1}), (\ref{rmm1}), (\ref{rmm2}) and
(\ref{he}). But, as is apparent, it differs only very little
from the value one, the maximal deviation of 12 percent
occurring at energies between 6 and 7 MeV, in backward
direction. The conclusion is that for such a reaction the
use of the approximation $T^C_{\gamma} \rightarrow
V^C_{\gamma}$ in (\ref{telg}) appears to be well justified.

Similarly to the atomic case, the Coulomb-Born approximation
quickly ceases to be acceptable if the charge of one of
the charged particles is increased. This is
illustrated in Fig. 11, where we show the ratio
$\left \vert R_{p\,C^{6+},0 0}({\rm{\bf{q}}}'_{\alpha},
{\rm{\bf{q}}}_{\alpha}; E) \right \vert$ for the reaction
of $C^{6+}$ colliding with deuterons (again, $C^{6+}$ is
considered a structureless nucleus). Here, the ratio
drops to 0.47 (at 12 MeV and a scattering angle of 67 degree).

\newpage
\section{Discussion} \label{disc}

We have investigated the quality of the replacement
of the two-particle Coulomb amplitude by the Coulomb potential
in the driving term for elastic and inelastic scattering,
both for atomic and nuclear three-body reactions. We have
found the following interesting results. \\
(i) For the atomic reactions studied, the Coulomb-Born
approximation quite generally turns out to be completely
unsatisfactory. Its failure becomes the more striking the
heavier the masses of the particles involved in the
intermediate-state Coulomb scattering are, and - not
surprisingly - the higher their charge is. \\
(ii) In the nuclear cases considered the situation is much
more favourable. At least for proton-deuteron scattering the
Coulomb-Born approximation, which on the energy shell has been
shown to be generally accurate to better than 10 per cent, is
fully justified (this is particularly so, since the
effective-potential part (\ref{telg}) itself represents
only a correction to the dominant driving term, which is
induced by the nuclear interaction; in fact, its Coulomb-Born
approximation (\ref{velg}) is known to contribute something
of the order of 10 per cent to the final
proton-deuteron observables, cf. Alt {\em et al} 1985).
Of course, also here its quality is diminished if the
charge of one of the particles is increased, but the
effect is not so dramatic as for the analogous atomic system.
\\
(iii) For atomic as well as nuclear elastic scattering
processes it has been found that the exact driving term
(\ref{telg}) becomes
equal to the approximate one, eq. (\ref{velg}), not only
in forward direction and for higher energies, but also
if the projectile energy goes to zero or, in other words,
if the elastic threshold is approached, for arbitrary
scattering angle. We have not yet succeeded in
developing a convincing argument for this unexpected fact.
A first suggestion, however, is based on the known invariance
of the ratio $\hat T^C_{\gamma}({\rm{\bf{p}}}',{\rm{\bf{p}}};
\hat E_{\gamma}+i0)/V^C_{\gamma}({\rm{\bf{p}}}',{\rm{\bf{p}}})$
under the transformations $p/k \to k/p, \quad p'/k \to k/p'$,
with $k = \sqrt{2 m_{\gamma} \hat E_{\gamma}}$ (cf, e.g., Kok
and van Haeringen 1980), which correlate high-energy
($k \to \infty$) with low-energy properties ($k \to 0$).
This speculation is supported by the fact that the value one
of the ratio (\ref{rel}) is reached for the larger and the
smaller projectile energies, respectively, the larger the
scattering angle becomes. \\
(iv) The minimum of the elastic ratio (\ref{rel}) is in all
cases attained for energies above the dissociation threshold,
frequently but not always in the backward direction. \\
(v) For excitation the Coulomb-Born approximation is
of similar quality as for the corresponding elastic reaction,
except for the additional failure in the near-forward
direction. Moreover, in this case the ratio (\ref{rel}) does
not approach the value one at the reaction threshold. \\
(vi) In all reactions considered the ratio (\ref{rel}), and
hence also the exact driving term (\ref{telg}), behave smoothly
when the bound state dissociation threshold is crossed. \\
(vii) In general, the imaginary part of the exact driving
term (\ref{telg}) is much smaller than its real part (the
Coulomb-Born approximation (\ref{velg}) is real everywhere),
except in the region of parameters where the ratio approaches
its minimum value. There they become comparable in magnitude.

{}From the numerical point of view it may be of interest to note
that the use of the Coulomb T-matrix in the calculation at
least of the on-shell driving term (\ref{telg}) for particles
with charges of equal sign, does not present any problem with
modern computers. Whether this holds true also for the calculation of
similar expressions but with oppositely charged particles,
or for the corresponding off-shell quantities (which would be
needed as input in the three-body integral equations approach)
remains to be seen.

\acknowledgments

\noindent
This work was supported by the Deutsche
Forschungsgemeinschaft,
Project no. 436 USB-113-1-0.\\

\newpage
{\bf References}

\noindent
Alt E O in Few Body Nuclear Physics (G. Pisent, V. Vanzani
and  L. Fonda, eds.), pp. 271, IAEA, Vienna (1978). \\
Alt E O, Grassberger P and Sandhas W Nucl. Phys. {\bf B2}, 167 (1967). \\
Alt E O, Sandhas W and Ziegelmann H Phys. Rev. {\bf C 17}, 1981 (1978). \\
Alt E O and Sandhas W Phys. Rev. {\bf C 21}, 1733 (1980). \\
Alt E O, Sandhas W and Ziegelmann H Nucl. Phys. {\bf A445}, 429 (1985);
Erratum Nucl. Phys. {\bf A465}, 755 (1987). \\
Alt E O, Avakov A R, Blokhintsev L D, Kadyrov A S and Mukhamedzhanov
A M J. Phys. B: At. Mol. Phys. {\bf 27}, 4653 (1994). \\
Chen J C Y and Chen A C Advan. At. Mol. Phys. 8, 71 (1972). \\
Faddeev L D Zh. Eksp. Teor. Fiz. {\bf 39}, 1459 (1960)
[Sov. Phys. -JETP {\bf 12}, 1014 (1961)]. \\
van Haeringen H Nucl. Phys. {\bf A327}, 77 (1979). \\
van Haeringen H {\em Charged-Particle-Interactions. Theory and
Formulas}, Coulomb Press, Leyden, 1985. \\
van Haeringen H and van Wageningen R J. Math. Phys. {\bf 16},
1141 (1975). \\
Kok L P, Struik D J, and van Haeringen H University of Groningen,
Internal Report 151 (1979). \\
Kok L P and van Haeringen H Phys. Rev. {\bf C 21}, 512 (1980). \\
Kok L P, Struik D J, Holwerda J E and van Haeringen H University
of Groningen, Internal Report 170 (1981). \\
Kok L P and van Haeringen H Czech. J. Phys. {\bf B32}, 311 (1982). \\
Kok L P and van Haeringen H Phys. J. Math. Phys. {\bf 25}, 3033 (1984). \\
Lewis R R Phys. Rev. {\bf 102}, 537 (1956). \\
Sinfailam A L and Chen Y C Phys. Rev. A{\bf 5}, 1218 (1972). \\
Veselova A M Teor. Mat. Fiz. {\bf 3}, 542 (1971). \\
Veselova A M Teor. Mat. Fiz. {\bf 35}, 395 (1978). \\

\newpage
{\bf Figure Captions}

\noindent
Fig. 1. Graphical representation of the elastic and inelastic
driving term ${\cal V}^{T^C}_{\gamma,n m}$. \\
Fig. 2. The integrand of ${\cal V}^{T^C}_{ee,00}
({\rm{\bf{q}}}'_{\alpha},{\rm{\bf{q}}}_{\alpha}; E)$
(cf. eq. (\ref{telg})) for $e + H(1s) \to e + H(1s)$ as a
function of the magnitude of the momentum k of the proton (the
integrations over all other variables are already performed)
at incident c.m. energy of 100 eV and at c.m. scattering
angle 20 deg.  Upper curve is the real part, and lower curve
the imaginary part of the integrand. Also shown is the
analogous integrand for the Coulomb-Born approximation
${\cal V}^{V^C}_{ee,00}({\rm{\bf{q}}}'_{\alpha},
{\rm{\bf{q}}}_{\alpha}; E)$, reduced by a factor of 5.
The arrow indicates the point where the $ee$-subsystem energy
is zero. The integrand and k are in a.u. \\
Fig. 3. The ratio $\vert R_{ee,00}({\rm{\bf{q}}}'_{\alpha},
{\rm{\bf{q}}}_{\alpha}; E) \vert$ as a function of c.m.
incident energy and c.m. scattering angle for elastic
electron + $H(1s)$ scattering. \\
Fig. 4. The ratio $\vert R_{p e^+,00}({\rm{\bf{q}}}'_{\alpha},
{\rm{\bf{q}}}_{\alpha}; E) \vert$ as a function of c.m.
incident energy and c.m. scattering angle for
elastic positron + $H(1s)$ scattering. \\
Fig. 5. The ratio $\vert R_{pp,00}({\rm{\bf{q}}}'_{\alpha},
{\rm{\bf{q}}}_{\alpha}; E) \vert$ as a function of c.m.
incident energy and c.m. scattering angle for elastic
proton + $H(1s)$ scattering. \\
Fig. 6. The ratio $\vert R_{p C^{6+},00}({\rm{\bf{q}}}'_{\alpha},
{\rm{\bf{q}}}_{\alpha}; E) \vert$ as a function of c.m.
incident energy and c.m. scattering angle for elastic
scattering of $C^{6+}$ off $H(1s)$. \\
Fig. 7. The ratio $\vert R_{ee,01}({\rm{\bf{q}}}'_{\alpha},
{\rm{\bf{q}}}_{\alpha}; E) \vert$ as a function of c.m.
incident energy and c.m. scattering angle for the inelastic
reaction $e + H(1s) \to e + H(2s)$. \\
Fig. 8. The amplitude ${\cal V}^{T^C}_{ee,01}$ as a function
of c.m. incident energy and c.m. scattering angle for the
inelastic reaction $e + H(1s) \to e + H(2s)$. \\
Fig. 9. The amplitude ${\cal V}^{T^C}_{pp,01}$ as a function
of c.m. incident energy and c.m. scattering angle for the
inelastic reaction $p + H(1s) \to p + H(2s)$. \\
Fig. 10. The ratio $\vert R_{pp,00}({\rm{\bf{q}}}'_{\alpha},
{\rm{\bf{q}}}_{\alpha}; E) \vert$ as a function of c.m.
incident energy and c.m. scattering angle for elastic
proton-deuteron scattering. \\
Fig. 11. The ratio $\vert R_{p C^{6+},00}({\rm{\bf{q}}}'_{\alpha},
{\rm{\bf{q}}}_{\alpha}; E) \vert$ as a function of c.m.
incident energy and c.m. scattering angle for elastic
$C^{6+} - d$ scattering. \\

\end{document}